\documentclass[fleqn,usenatbib]{mnras}

\usepackage{newtxtext,newtxmath}
\usepackage{aas_macros}
\usepackage{aas_macros}

\usepackage{paralist}
\usepackage{amsmath}
\usepackage{xspace}
\newcommand{\eagle}{\textsc{eagle}\xspace}
\newcommand{\lya}{{\hbox{\rm Lyman-$\alpha$}}}
\newcommand{\teff}{\ensuremath{\tau_{\rm eff}}}
\newcommand{\wdm}{{\text{\sc wdm}}} 
\newcommand{\dm}{{\text{\sc dm}}} 
\newcommand{\eff}{{\text{\rm eff}}}

\usepackage{graphicx}
\graphicspath{{fig/}} 
\usepackage{paralist}
\usepackage{longtable}
\usepackage{units} 
\usepackage[dvipsnames,usenames]{color,xcolor}
 
\usepackage{soul}
\definecolor{antiquewhite}{rgb}{0.98, 0.92, 0.84}
\sethlcolor{antiquewhite}
\usepackage[obeyFinal,obeyDraft,textsize=scriptsize,textsize=footnotesize,color=antiquewhite]{todonotes}

\newcommand{\myparagraph}[1]{\par\bigskip\noindent\textbf{#1}}

\title[How warm is too warm?]{How to constrain warm dark matter with the Lyman~$\alpha$ forest}

\author[A. Garzilli et al.]{
Antonella Garzilli,$^{1,2}$\thanks{E-mail: garzilli@nbi.ku.dk}
Andrii Magalich,$^{3}$
Oleg Ruchayskiy$^{2}$
and Alexey Boyarsky$^{3}$\\
$^{1}$EPFL Laboratoire d'astrophysique, Observatoire de Sauverny, CH-1290 Versoix, Switzerland\\
$^{2}$Niels Bohr Institute, Copenhagen
  University, Blegdamsvej 17, DK-2100 Copenhagen, Denmark\\
$^{3}$Lorentz Institute, Leiden University, Niels Bohrweg 2,
  Leiden, NL-2333 CA, The Netherlands
}

\date{Accepted 2021 January 20. Received 2021 January 20; in original
  form 2019 December 19}

\pubyear{2021}

\begin{document}
\label{firstpage}
\pagerange{\pageref{firstpage}--\pageref{lastpage}}
\maketitle

\begin{abstract}
  The flux power spectrum of the high resolution Lyman-$\alpha$ forest data 
  exhibits suppression at small scales.
  The origin of this suppression can be due to long-sought warm dark matter (WDM) or to thermal effects, related to the largely unknown reionization history of the Universe.
  Previous works explored a specific class of reionization histories that exhibit sufficiently strong thermal supression and leave little room for warm dark matter interpretation.
  In this work we choose a different class of reionization histories,
  fully compatible with available data on evolution of reionization,
  but much colder then the reionization histories used by previous
  authors in determining the nature of dark matter, thus leaving the broadest room for the WDM interpretation of the suppression in the flux power spectrum.
  We  find that  WDM thermal relics with masses below 1.9~keV (95\% CL) would produce a suppression at scales that are larger than observed maximum of the flux power spectrum, independently of assumptions about thermal effects.
  This WDM mass is significantly lower than previously claimed bounds,
  demonstrating the level of systematic uncertainty of the
  Lyman-$\alpha$ forest method, due to the previous modelling.
  We also discuss how this uncertainty may affect also data at large scales measured by eBOSS.
\end{abstract}

\section{Warm dark matter and Lyman-$\alpha$ forest}
\label{sec:intro}

Dark matter in the Universe manifests itself through many distinct observations -- from rotational curves of stars in galaxies to the statistics of anisotropies of cosmic microwave background~\cite{Peebles:2017bzw}.
If dark matter is made of particles -- little is known about their properties.
In particular, it is not known whether dark matter is \textit{warm} -- \emph{i.e.} was born relativistic -- as opposed to \textit{cold dark matter} (CDM).
While warm dark matter (WDM) was cooling with expansion, it erased primordial inhomogeneities at scales below its \emph{free streaming} horizon, $\lambda_\dm$. This means that the growth of structures in warm dark matter is  suppressed.
The main distinction between CDM and WDM lies thus in the suppression of the matter power spectrum below a certain scale.

One of the promising tools to measure dark matter distribution at small scales is the \emph{Lyman-$\alpha$ forest method} -- a set of absorption features in the spectra of background quasars due to the Lyman-$\alpha$ ($n=1\rightarrow 2$) transition in the rest frame of neutral hydrogen clouds at redshifts $2 \le z \le 5$ \citep{meiksin2009}.
Under the assumption that neutral hydrogen traces matter distribution, it allows to measure a proxy of the matter power spectrum (known as \emph{flux power spectrum}, FPS).
The method has been actively used to explore the properties of  dark matter particles~\citep{Hansen01,Viel05,Viel06,Seljak06,Viel:2007mv,Boyarsky09a,Boyarsky:2008mt,Viel13,Garzilli:2015iwa,irsic2017,Murgia:2018now,Baur16,baur2017,Garzilli18,Palanque-Delabrouille:2019iyz}.

The effect of warm dark matter gets more pronounced at smaller scales.
Therefore, Lyman-$\alpha$ forest data based on spectra of high spectral resolution ($\mathcal{R} \equiv \lambda/\Delta \lambda \sim 60'000$) are of special interest as they allow to probe FPS down to scales $k\sim \unit[0.1-0.2]{s/km}$, see \citet{Garzilli18}. 
The observed flux power spectrum $\Delta^2_F $ exhibits a sharp suppression at comoving wave-number above $k_F\sim \unit[0.03]{s/km}$~\citep{Boera:2018vzq}.
The same cutoff is visible in the older dataset based on HIRES~\citep{Vogt:1995zz} and MIKE~\citep{MIKE} quasar spectra \citep{Viel13}.

\subsection{Degeneracy between WDM and thermal effects}
The presence of the cut-off in the high resolution data does not mean, however, that warm dark matter with $\lambda_\dm \sim \lambda_F$ has been discovered!
Indeed, several thermal effects may prevent {\sc hi} from following dark matter distribution at small scales.
These effects are  \emph{degenerate} with that of warm darm matter.
This makes the derivation of robust WDM bounds from the high resolution Lyman-$\alpha$ data problematic, as we will discuss in details below.

\myparagraph{Doppler broadening.} Intergalactic medium (IGM) is heated during reionization.
The temperature of the intergalactic medium depends on the density and follows the \emph{temperature-density relation} \citep{hui1999}:
\begin{equation}
        \label{eq:TDR}
        T = T_0(z) \left(\frac{\rho}{\bar \rho}\right)^{\gamma(z) -1}
\end{equation}
where $\rho$ is the matter density of a small patch of the Universe and $\bar\rho$ is the background matter density.
Both parameters $T_0$ and $\gamma$ are functions of redshift.
The non-zero temperature and thus the Maxwellian velocities of the gas particles cause the  Doppler broadening of the absorption lines. This introduces its own cutoff in the flux power spectrum.
The temperature of the gas,  and hence the scale of the Doppler (temperature) broadening, $\lambda_b$ is not accurately known \citep[see e.g.][]{Garzilli:2015iwa, Rorai18}, especially at redshifts $z\gtrsim 5$ \citep{hui2003,bolton2010,becker2011,bolton2012,Lidz:2014jxa,Walther:2017cir,Walther:2018pnn}.
Therefore, in deriving WDM bounds one marginalizes over $T_0(z)$ and $\gamma(z)$.
 
The  high resolution HIRES/MIKE data were used to derive Lyman-$\alpha$ constraints by \citet{Viel13}. Assuming a particular family of reionization histories \citep{haardt2001,Haardt12} combined with a powerlaw ansatz for the $T_0(z)$ for $3 \le z \le 6$, the constrain $m_\wdm \ge 3.3$~keV was found.\footnote{In this work by ``WDM'' we refer to a specific class where DM is produced in thermal equilibrium, and then freezes out -- \textit{(warm) thermal relics} or ``thermal WDM''~\protect\citep{Bode:2000gq}.
In this case there is a one-to-one relation between {$\lambda_{\dm}$} and the DM particle mass, $m_{\wdm}$ ({$\lambda_{\dm} \propto
m_\wdm^{-4/3}$}), \citep[see e.g.][]{Boyarsky:2008mt}.
Although ``thermal WDM'' does not correspond to any specific particle physics model, such a parametrization of WDM is often chosen when deriving constraints. 
We use it in our paper to faciliate comparison with other works.}
This would correspond to the cutoff scale $\lambda_{\rm dm} \lesssim
30$~ckpc.

These bounds were revisited in~\citep{Garzilli:2015iwa}.
It was argued that the IGM temperature is poorly constrained at redshifts $z\sim 5$ and may not be monotonous (due to combination of adiabatic cooling and HeII reionization starting at later times).
By allowing $T_0(z)$ to be non-monotonous at $z = 4-6$, 
\citet{Garzilli:2015iwa} demonstrated that the data can be fit by the
WDM model with $m_\wdm \simeq 2.1$~keV or by the CDM (\textit{i.e.}
the \emph{two models could not be distinguished by the data}).
The fact that their bounds weaken if $T_0(z)$ is allowed to vary independently in each redshift bin has been acknowledged  by \citet{Viel13} who did not find, however, statistical evidence in favour of it and chose to quote a less conservative WDM bound, see \citet{Garzilli:2015iwa} for discussion.

Re-analysis of the same data with higher resolution of simulations,
different modeling of ultra-violet background~\citep{irsic2017}
changed the bound to around $m_\wdm \ge 3.9-4.1$~keV or confirmed our
findings, depending on the choice of priors on thermal history. 

Later works \citep{Walther:2017cir,Walther:2018pnn,Boera:2018vzq} indeed confirmed that IGM temperature reaches a local minimum around $z\sim 5$.

\myparagraph{Pressure effects.} The proper inclusion of the temperature effects did not however, exhausts all possible uncertainties of the WDM bounds.
Inceed, as the absorbing filaments in the IGM are not virialize  structures, their typical size \emph{depends on the past  thermal history}, relatively to the time of observation.  The gas  distribution is smoothed compared to the dark matter due to pressure  \citep{Gnedin98, Theuns00,Kulkarni:2015fga,Upton16}

In the absence of  realistic first-principle modeling of reionization, one has to run hydrodynamical simulations with pre-defined photo-heating and photo-ionization rates in order to assess the influence of each specific thermal history onto the Lyman-$\alpha$ observables.
Properly quantifying systematic effects of all admissible reionization histories would require running a prohibitively large grid  of hydrodynamical simulations.

The work \citet{Garzilli18}  demonstrated that the observed shape of the high-resolution FPS and its redshift evolution can be explained equally well by either CDM plus \citet{Haardt12} reionization history, or by warm dark matter plus a reionization history of \citet{onorbe2016} that has a lower scale $\lambda_p$ still perfectly consistent with the data.
The work did not put specific bounds, but rather illustrated that the \emph{degeneracy between astrophysical and warm dark matter is much stronger than assumed in the previous works}.
This degeneracy cannot be resolved by the current data (although many interesting ideas exist on how this can be done \citep[see e.g.][]{Garzilli:2015bha,Garzilli:2018jna,Gaikwad:2020art}

Until this is done -- \emph{unknown reionization history represents the major systematic uncertainty in deriving bounds on WDM mass}.

Current work makes a new step in evaluating these uncertainties.

In order to quantify the uncertainty, we find a  reionization history with the smallest pressure support scale, $\lambda_p$ \citep{onorbe2016}. 
By pushing all astrophysical effects to their minimum, we try to describe the cut-off in the data by the warmest possible DM model.
In this way, we identify the WDM model whose cut-off scale is larger than $\lambda_F \sim \unit[30]{(s/km)^{-1}}$ and is, therefore, ruled out independently of astrophysical assumptions.
Therefore, when allowing for the wider class of reionization histories, the WDM bound relaxes down to $1.9$~keV (which represents a factor of $\approx 3$ increase of the characteristic cut-off scale, $\lambda_\dm \lesssim 100$~ckpc).
This results is defined by the scale of the cut-off observed in the data, the bound can not be improved unless  the origin of the cut-off is proven to be astrophysical.

\myparagraph{Other WDM bounds.}
If the bounds from high-redshift data indeed have large uncertainty -- can  other data provides more robust bounds? 
The  Lyman-$\alpha$ data with medium spectral resolution ($\mathcal{R} \sim 2'200 $), coming \textit{e.g.}\ from the SDSS (BOSS, eBOSS) surveys has been used to constrain WDM \citep{Viel06,Seljak06,Viel:2007mv,Boyarsky09a,Boyarsky:2008mt,Baur16,baur2017,Palanque-Delabrouille:2019iyz}.
These data allow to probe comoving wave-numbers up to $k_{\rm sdss} \simeq \unit[0.02]{sec/km}$.
The most recent bound using the SDSS DR14 (eBOSS) data was claimed to be $m_\wdm \gtrsim 5.3$~keV at 95\%CL \citep{Palanque-Delabrouille:2019iyz}.
 Both temperature and pressure effects operate at scales smaller than $k_{\rm sdss}^{-1}$.
\citet{Palanque-Delabrouille:2019iyz}  limits WDM by requiring that its deviation from CDM (inferred from large-scale cosmological observations) at scales above the cut-off ($k< k_{\rm sdss}$) would be ``within errorbars".
Statistical errorbars of the eBOSS data are small, owing to large number of observed quasars and the bound based on these statistical errors appear to be quite tight. 
We arrive to a paradoxical conclusion that the scales where WDM effects are small (at $k < k_{\rm sdss} < k_F$) are more sensitive than the scales where they are large ($k > k_F$, high-resolution data). This conclusion relies however on the fact that the sensitivity of the eBOSS data
is defined by statistical rather than systematic uncertainties. To make this conclusion one 
needs to model astrophysical effects in great detail. 
In particular, if pressure effects and effects of warm dark matter are of the same order at scales $k > k_F$ \citep{Garzilli18}, one should demonstrate that their $k$ dependence is different. 
Otherwise, two models: ``CDM plus larger $\lambda_p$'' and ``WDM plus smaller $\lambda_p$'' that were indistinguishable at $k > k_F$ will be indistinguishable also at $k < k_{\rm sdss}$.
\citet{Palanque-Delabrouille:2019iyz} and previous works \citep{Viel06,Seljak06,Viel:2007mv,Boyarsky09a,Boyarsky:2008mt,Baur16,baur2017} did not perform such an analysis. 
In particular, they did not explore in their work the full range of possible reionization histories, including ``Late'' versions of \citep{onorbe2016,onorbe2017}.
They limited their redshift of reionization to $z_{\rm reio} > 7$, while the recent data suggests that hydrogen reionization might have started at $z_{\rm reio} < 7$~\citep{onorbe2016,onorbe2017,Walther:2018pnn,Boera:2018vzq}) and was completed by $z=5.7$ \citep{Becker:2014oga,schroeder2013,Becker:2001ee}.
We see therefore, that although formally stronger, \emph{bounds based on eBOSS data are not free from systematic uncertainties due to unknown reionization history}.

\bigskip

Below we derive the bounds on the WDM particle
mass using the high resolution Lyman-$\alpha$ data and the reionization history from \citep{onorbe2016} with the smallest pressure effects. 
We use the high-resolution data from \citet{Boera:2018vzq}. 
This dataset is a re-observation with longer exposure of the quasar spectra already covered by the dataset of  \citep{Viel13} (for their comparison see Appendix~C of \citet{Boera:2018vzq}).
We find that WDM models with $1.9$~keV  have cut-off at scales around maximum of FPS ($k\sim k_F \approx \unit[0.03]{s/km}$) and are therefore ruled out  independently of astrophysical assumptions.

\begin{table*}
\centering
\begin{tabular}{llllc}
  \hline
  Name & $L\, [{\rm Mpc}/h]$ & $N$ & Dark matter & {UVB}\\
  \hline
  CDM & 20 & $1024^3$ & CDM & {\sc LateCold}   \\
  WDM & same & same & $m_{\rm WDM} = 2\,{\rm keV}$ & same \\
  \hline
  EAGLE\_REF & 100 $ h$ & $1504^3$ & CDM & {\sc Eagle}\\
  \hline
\end{tabular}
\caption{\label{tab:simulations} Hydrodynamical simulations considered
  in this work together with corresponding parameters. All simulations
  were performed specifically for this work, except {\tt EAGLE\_REF}
  \citep{schaye2015}.  Columns contain from left to right: simulation
  identifier, co-moving linear extent of the simulated volume ($L$),
  number of dark matter particles ($N$; also equal to the number of
  gas particles), type of dark matter (CDM or WDM), $m_{\rm WDM}$ (expressed in natural units),
  ultra-violet background imposed during the simulation
  ({\sc LateCold} refers to a reionization model in
  agreement with measured temperature of the IGM as discussed in
  \protect\citet{onorbe2016}, see Fig.~\ref{fig:LateCold};
  {\sc Eagle} indicates the standard UVB from \citep{haardt2001}).
  The cosmological parameters are chosen according to
  Table~\protect\ref{tab:cosmology}. The gravitational softening
  length for gas and dark matter is kept constant in co-moving
  coordinates at 1/30$^{\rm th}$ of the initial interparticle
  spacing. All simulations start from the initial conditions
  generated by the \texttt{2LPTic} \protect\citep{scoccimarro2012}
  with the same \lq glass\rq-like particle distribution generated by
  GADGET-2 \protect\citep{springel2005}.}
\label{tab:sims}
\end{table*}

\begin{table}
\centering
\begin{tabular}{lll}
\hline
Cosmology & \emph{Planck} \citep{Planck2015} \\
\hline
$\Omega_0$       & $0.308 \pm 0.012$   \\
$\Omega_\Lambda$ & $0.692 \pm 0.012$   \\
$\Omega_b h^2$   & $ 0.02226 \pm 0.00023$ \\
$h$              & $0.6781 \pm 0.0092$  \\
$n_s$            & $0.9677 \pm 0.0060$  \\
$\sigma_8$       & $0.8149 \pm 0.0093$  \\
\hline
\end{tabular}
\caption{\label{tab:cosmology}Cosmological parameters used in our
  simulations. \emph{Planck} cosmology is the conservative choice of
  TT+lowP+lensing from \citet{Planck2015} (errors represent $68\%$
  confidence intervals).}
\label{tab:cosmo}
\end{table}

\begin{figure}
    \centering
    \includegraphics[width=\linewidth]{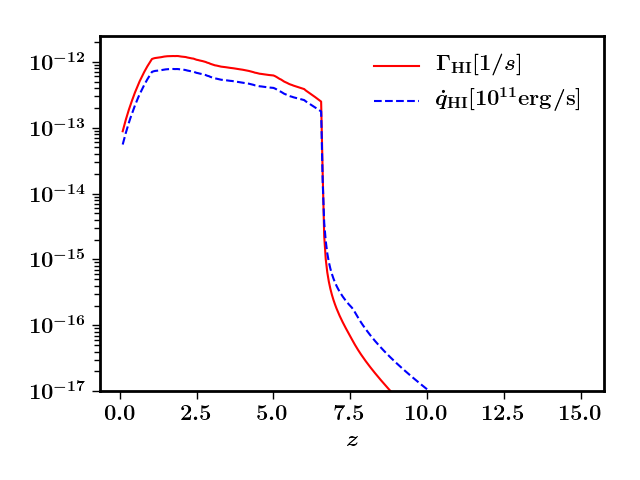}
    \includegraphics[width=\linewidth]{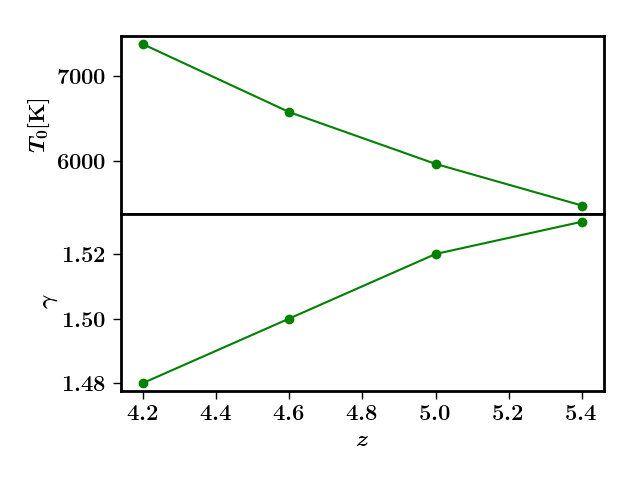}
    \caption{The adopted thermal history. In the upper panel we show the photoionization ($\Gamma_{\text{\sc hi}}$) and photoheating ($\dot q_{\text{\sc hi}}$) rates for the neutral hydrogen ({\sc hi}) from the {\sc LateCold} model of~\protect\citet{onorbe2016}. The photoheating rate is rescaled by $10^{11}$ in order to fit onto the same plot.
    The reionization in this model starts at $z_{\rm reio}= 6.7$. In the lower panel we show the resulting temperature at the cosmic mean density of the IGM, $T_0$, as well as the slope of the temperature-density relation, $\gamma(z)$ (obtained by fitting the Eq.~\eqref{eq:TDR} to the simulation snapshots).}
    \label{fig:LateCold}
\end{figure}

\section{Our method} 
We start with a reionization history that provides smallest pressure support ($\lambda_p \ll \lambda_F$) and marginalize over the Doppler broadening and over the $\teff$.
In this way we arrive to a model where \emph{all astrophysical effects are pushed to the minimal level} and thus there is the broadest room for WDM effect. In more details our procedure is the following:
  \begin{asparaenum}[\bf (1)]
\item 
    In order to minimise the pressure effects, we  run hydrodynamical 
    simulations using the  model of ultraviolet background (UVB) 
    {\sc LateCold}  \citep{onorbe2016}.\footnote{The details about the
      numerical simulations can be found in \citep{Garzilli18}.} We
    have summarized the
    simulations in Table~\ref{tab:sims}, with cosmological parameters
    listed in Table~\ref{tab:cosmo}.
    In this model the reionization starts at redshift $z=6.7$,
    later than other thermal histories, considered in~\cite{onorbe2016} or by other groups. Nevertheless, {\sc LateCold} scenario  reproduces the measured temperature at post-reionization redshift $z \sim 5$  compatible with the constraint on reionization time~\cite{onorbe2016}.
    By construction it gives the minimal filaments size.\footnote{We thank J.~Onorbe for sharing with us the data of the {\sc LateCold} thermal history that were not published together with the other thermal histories in~\cite{onorbe2016}.} {\sc hi} photo-heating and photo-ionization rates are shown in Fig.~\ref{fig:LateCold}. We have assumed that the intergalactic medium is optically thin, and that reionization happens uniformly in all the space.
  \item We explicitly marginalise over the IGM temperature (\textit{Doppler broadening}) adding their effects to the FPS in post-processing.
    Because the Lyman~$\alpha$ forest at each redshift is sensitive only to a short range of densities, we model the IGM temperature with a single value, $T_0$ of the temperature-density relation~\eqref{eq:TDR}.  This is additionally justified by the fact that in the approximation of instantaneous reionization, that we are using, all the part of the IGM get to the same temperature at the same time, hence $\gamma=1.0$, \textit{i.e.}, there is no dependence of temperature on density.

  \item We also marginalise over the \emph{effective optical depth}
    $\tau_\eff =-\ln\langle F \rangle$, where $F$ is the transmitted
    flux, and $\langle\ldots\rangle$ is the average.  $\tau_\eff$
    encompasses the information about the average absorption level
    (\emph{i.e.}  overall level of ionization of the IGM).  The
    work~\cite{Boera:2018vzq} provided measurements of $\tau_\eff$ in
    each of the redshift bins together with the errorbars.  We can
    vary $\tau_{\rm eff}$ in the post-processing of the spectra, by
    rescaling the optical depth in the spectra by a suitable factor.
    \end{asparaenum}

\begin{figure*}
  \centering \includegraphics[width=0.33\textwidth]{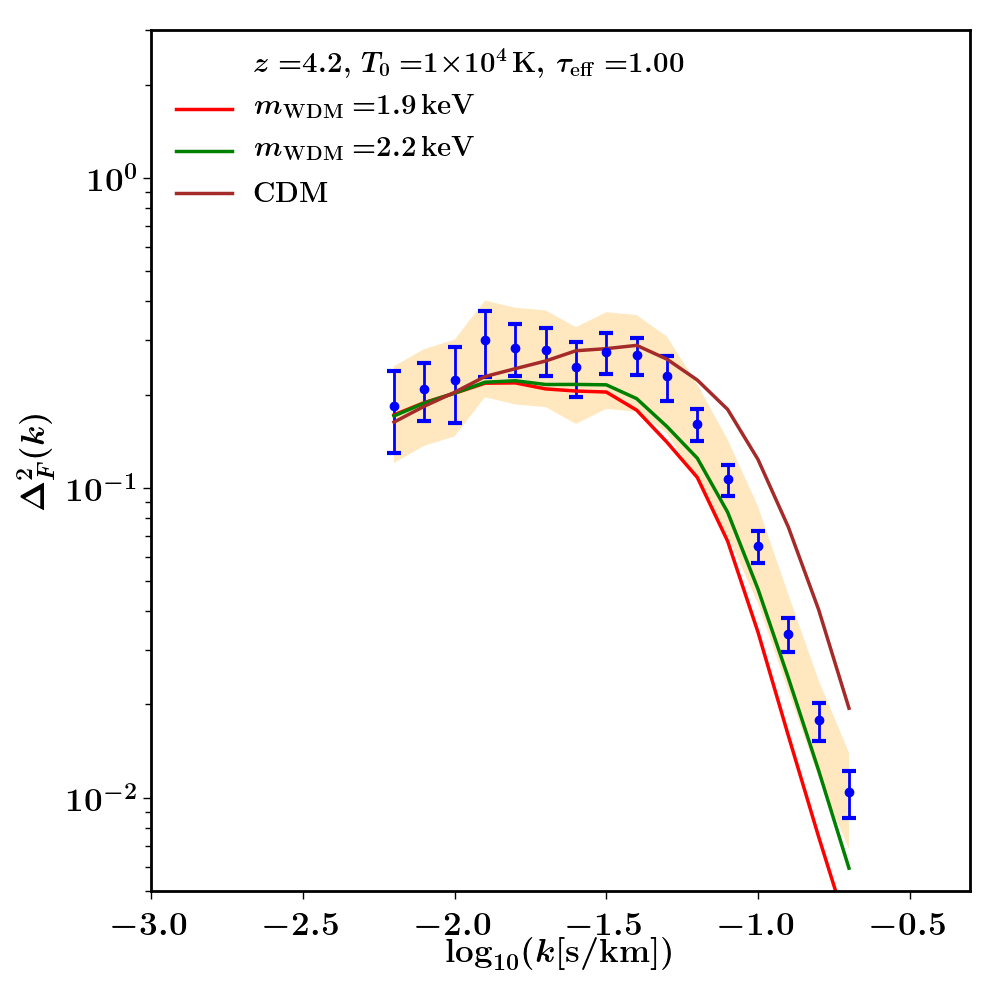}~\includegraphics[width=0.33\textwidth]{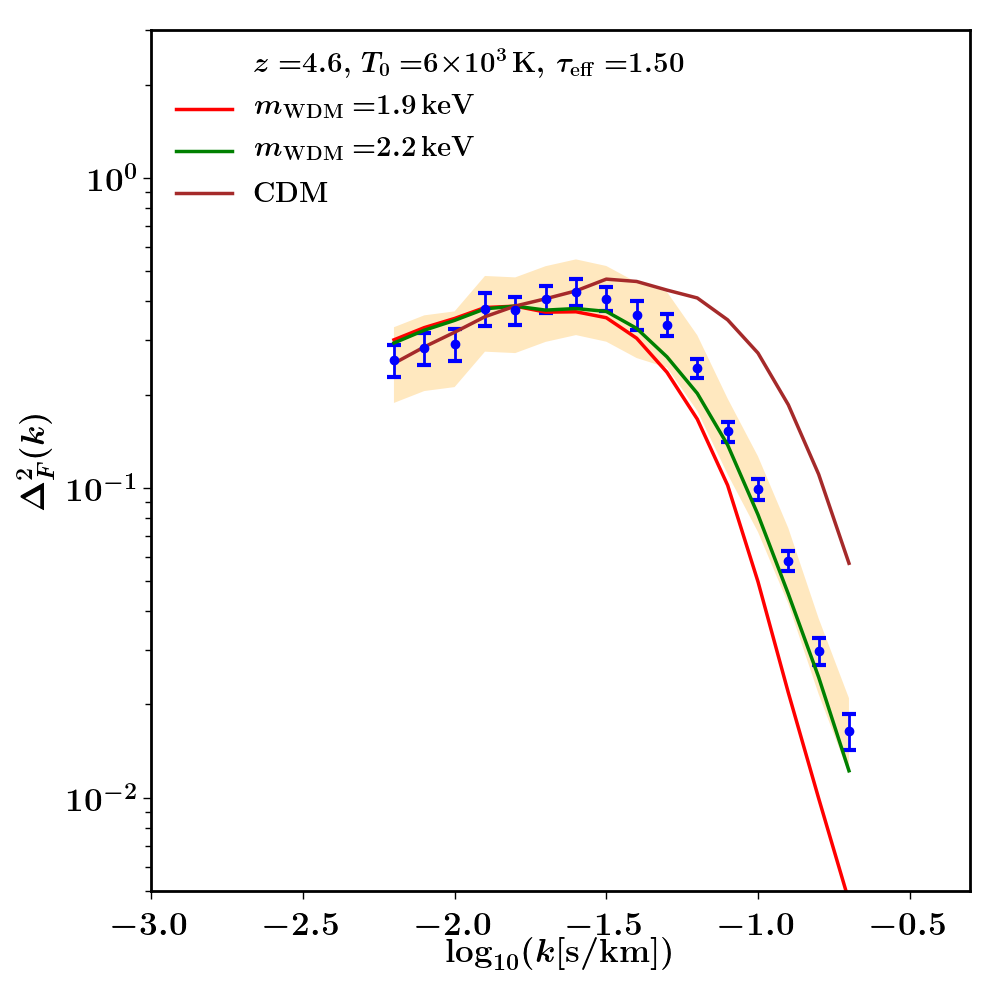}~\includegraphics[width=0.33\textwidth]{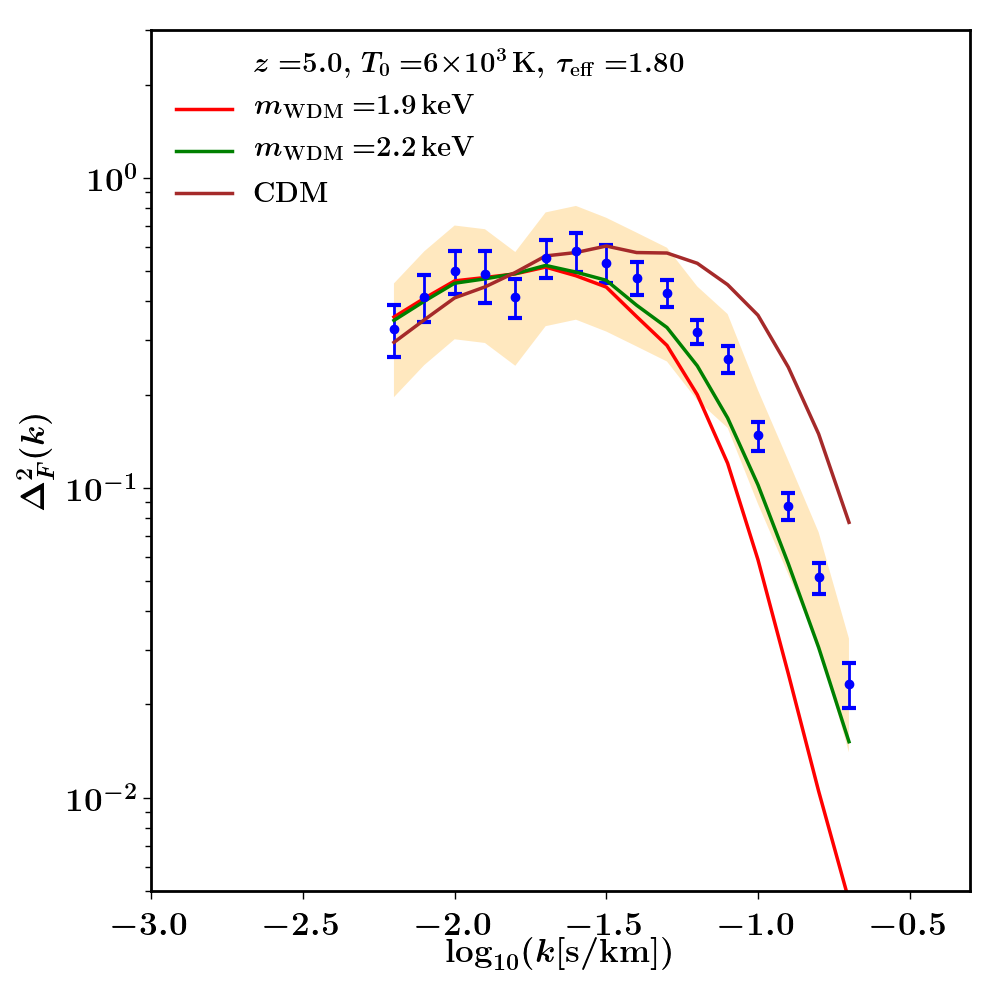}
  \caption[]{Data of \protect\citeauthor{Boera:2018vzq} together with several cosmological  models: WDM  with $m_{\wdm}=1.9\,{\rm keV}$ (red line, compatible with the data at $95\%$~CL), WDM with $m_\wdm = 2.2$~keV (green line) and CDM (brown line).
  The thermal history {\sc LateCold} from \cite{onorbe2016} with late redshift of reionization $z_{\rm reio} = 6.7$ is used for all three models.
        Orange shaded region indicates the level of uncertainty of the data, related to the sample variance (estimated from the simulations, see text for details).}
  \label{fig:figPS}
\end{figure*}

\subsection{Data and Covariance matrix}
The HIRES sample of high
resolution quasar spectra at $z>4$, released by \citet{Boera:2018vzq},
consists of 15 Lyman~$\alpha$ forest spectra over the redshift range
$4.0\leq z \leq 5.2$.   
These spectra
are binned into three redshift intervals of width $\Delta z =0.4$,
centered on $z=4.2,4.6,5.0$.  Their total comoving length per redshift
interval is $L=1412,2501,1307\,{\rm Mpc}/h$.  By construction the
\lya\ data points are strongly correlated~\cite{rollinde2013}.  The
covariance matrices have been estimated in \cite{Boera:2018vzq} from
the data, using ``bootstrap method''.  While bootstrap method is
widely used in the literature for estimating the covariance matrix, it
can only evaluate covariance due to the scatter within the sample (see
\textit{e.g.},~\cite{rollinde2013}).  From the data one cannot
estimate, however, how representative the sample is -- this
contribution is also commonly called \textit{sample variance}. Because
of the smallness of the sample size the sample variance can be an
important contribution to the covariance matrix. We try to estimate it
using the simulations.  To properly recover a covariance matrix from
the simulations, one needs a simulation box whose size is comparable
with the comoving length of the spectra (see \textit{e.g.}, discussion
in \citet{Garzilli18}).  A typical length of a single Lyman-$\alpha$
forest spectrum is $\sim 200\,{\rm cMpc}$, while our simulation box is
$\unit[20]{cMpc}$.

In order to evaluate the covariance matrix  we took the {\eagle}
simulation  with $L=100\,{\rm Mpc}/h$ and $N_{\rm part}=1504^3$
\citep{schaye2015}.  These simulations do not have the
resolution needed for the smallest scales measured in the dataset ~\citep[see
the discussion in][]{Garzilli18}.
For this reason, we  approximate the covariance matrix with the following
procedure.
We computed the expected covariance matrix from
the above mentioned \eagle simulation run by bootstrap, considering a
mock data sample of the length equivalent to that of the observed data
sample. For each redshift interval, we consider the snapshot at the
central redshift. We extract 1000 mock spectra. We compute the FPS for
each mock spectrum. We group the mock spectra in sub-samples with the
same comoving length as the data-sample, and for each mock sub-sample
we compute the average of the FPS. The covariance matrix is computed
with the average FPS with respect to the average FPS computed on all
the mock spectra.  After that, we compute the maximum relative
  errors obtained by comparing the diagonal elements of the covariance
  matrix to the data.  Hence, we rescale all elements of the
  covariance matrix to this same maximum relative error.
Our normalized matrices does not coincide with those of
\citep{Boera:2018vzq}, but their diagonal terms are comparable in magnitude, the difference can be attributed to a different resolution. Indeed, we checked that insufficient resolution \textit{increases} a correlation between data points.

In Figure~\ref{fig:figPS} we show a direct comparison between the data
and three distinct cosmologies  with LateCold thermal history. While
the WDM cosmology with $m_{\rm WDM}=2.2$~keV is excluded at
2-$\sigma$, the WDM cosmology with $m_{\rm WDM}=1.9$~keV is compatible
with the data at 95\% CL, and the the CDM cosmology is compatible with
the data. The reason is that we claim that all the models whose flux PS lies
below the observed flux PS must be excluded in our analysis. In fact,
among all the thermal histories that are compatible with the
observations, we have considered the thermal history that provide the
minimal filtering length. Hence a model whose flux PS lies below the
observations cannot become compatible with the data with another
thermal history. Instead a model whose flux PS is above the
observations may be compatible with the data if another thermal
history is considered.

\subsection{Data analysis} Our resulting model describing evolution of FPS contains 7 parameters: inverse WDM mass $[\unit{keV}]/m_{\rm WDM}$, IGM temperature at the cosmic mean density and $\tau_{\rm eff}$ (the latter two quantities are evaluated at redshifts $z=4.2, 4.6, 5.0$).
We consider linear priors on the temperature and logarithmic priors on the mass of warm dark matter.
In order to get theoretical predictions of the FPS, we run our simulations for two distinct cosmological models: CDM, and thermal relic WDM with mass $\unit[2]{keV}$.
Starting from our simulations, we compute in post-processing the FPS
in $k$-space for each set of the parameters, the parameters are three
and they are arranged on a three-dimensional regular grid.
Our final theoretical model is obtained by interpolating linearly the FPS across the grid.
We perform a joint analysis on all the redshift intervals.

\section{Results}
\label{sec:results}

\begin{figure*}
  \centering 
  \includegraphics[width=0.65\columnwidth]{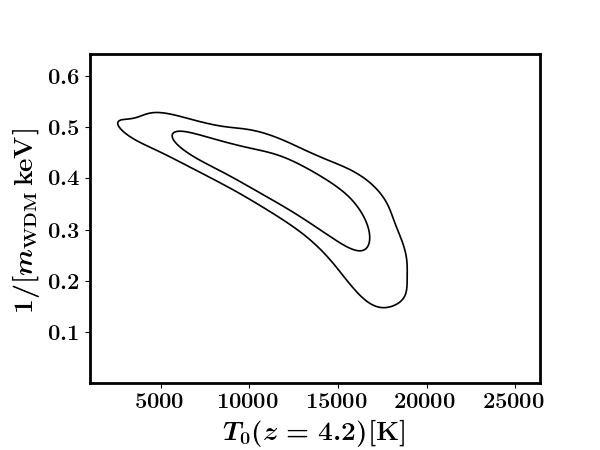}
  \includegraphics[width=0.65\columnwidth]{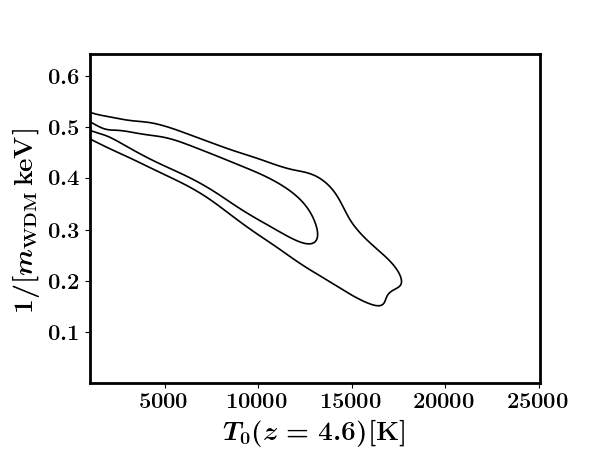}
  \includegraphics[width=0.65\columnwidth]{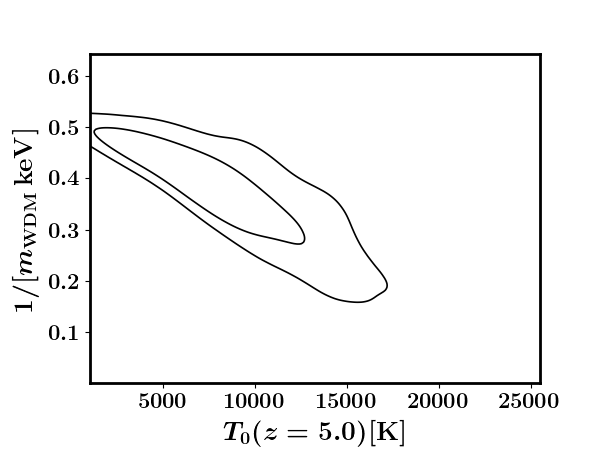}
  \includegraphics[width=0.65\columnwidth]{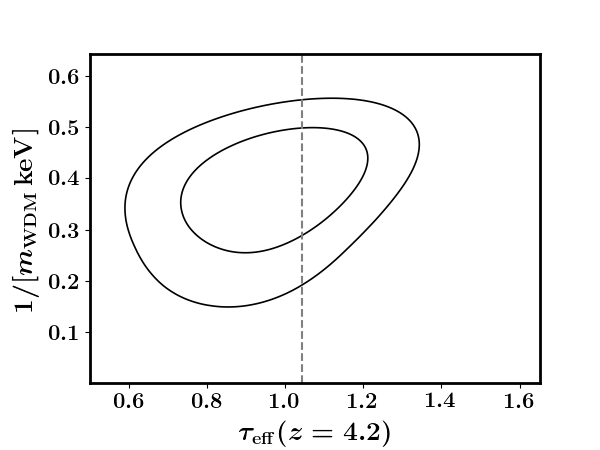}
  \includegraphics[width=0.65\columnwidth]{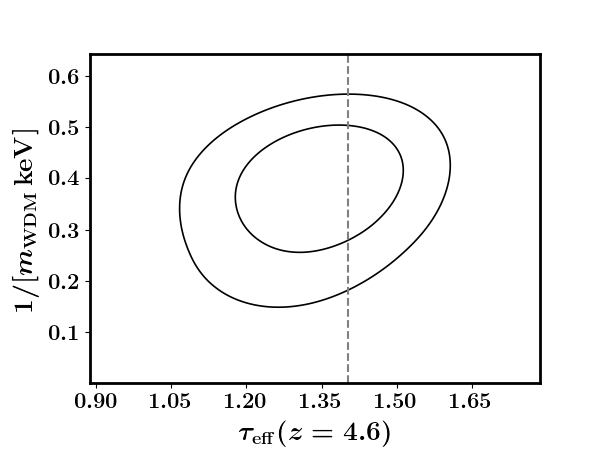}
  \includegraphics[width=0.65\columnwidth]{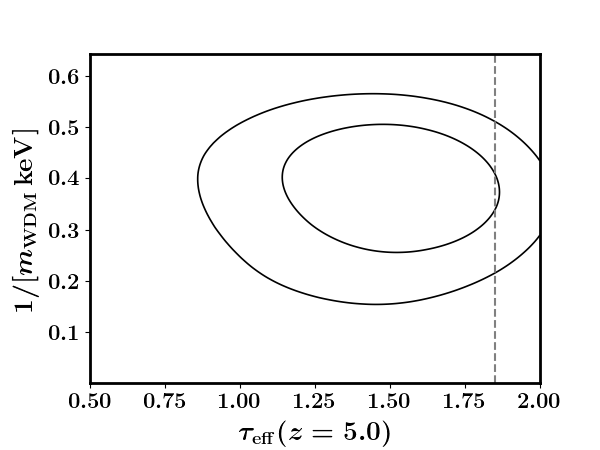}
  \caption[Confidence regions between the WDM mass, the IGM mean
    temperature and effective optical depth]{Confidence regions
    between the WDM mass, $m_\wdm$, the IGM mean temperature, $T_0$
    and the effective optical depth $\tau_\eff$ at redshifts $z=4.2,\,
    4.6, \, 5.0$, the $T_0$ upper bound limits are still dependent on
the specific choice of the pressure smoothing model.
    Our analysis shows that if {\sc LateCold} were a true history of reionization, then the CDM would be ruled out. However, it is not possible to use this analysis to determine the IGM temperature in CDM for reionization histories outside {\sc LateCold}. 
    For the same reason our analysis does not allow to determine robust \textit{lower} bounds on $T_0$ for a given WDM mass. The (conservative) upper bounds are shown in Fig.~\protect\ref{fig:t0_imass_mcmc}.}
  \label{fig:2D}
\end{figure*}

\begin{figure}
  \centering \includegraphics[width=\columnwidth]{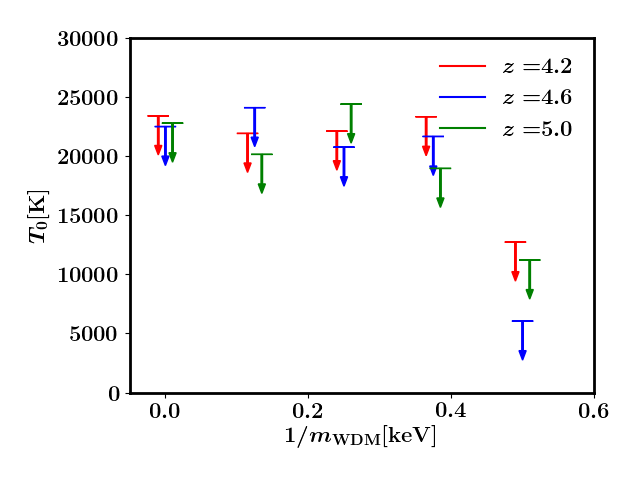}
  \caption[]{The 2-$\sigma$ level upper limit on $T_0$ as a function of $1/m_{\rm WDM}$. The upper limit has been estimated separately for each redshift interval with fixed $m_{\rm WDM}$. We only show the mass interval between $m_{\rm WDM}=\unit[2]{keV}$ and CDM. This result is in substantial agreement with the global fit that we have shown in Figure~\ref{fig:2D}.  These bounds are fully consistent with existing estimates of the IGM temperatures at redshifts $z=4-5$~\cite[see \textit{e.g.},][]{Schaye00,becker2007,becker2011,bolton2012}.
  One should keep in mind that these works evaluated IGM temperatures for CDM cosmology only and for some limited class of thermal histories, not necessarily consistent with {\sc LateCold}.}
  \label{fig:t0_imass_mcmc}
\end{figure}

\begin{figure*}[!t]
  \centering \includegraphics[width=\textwidth]{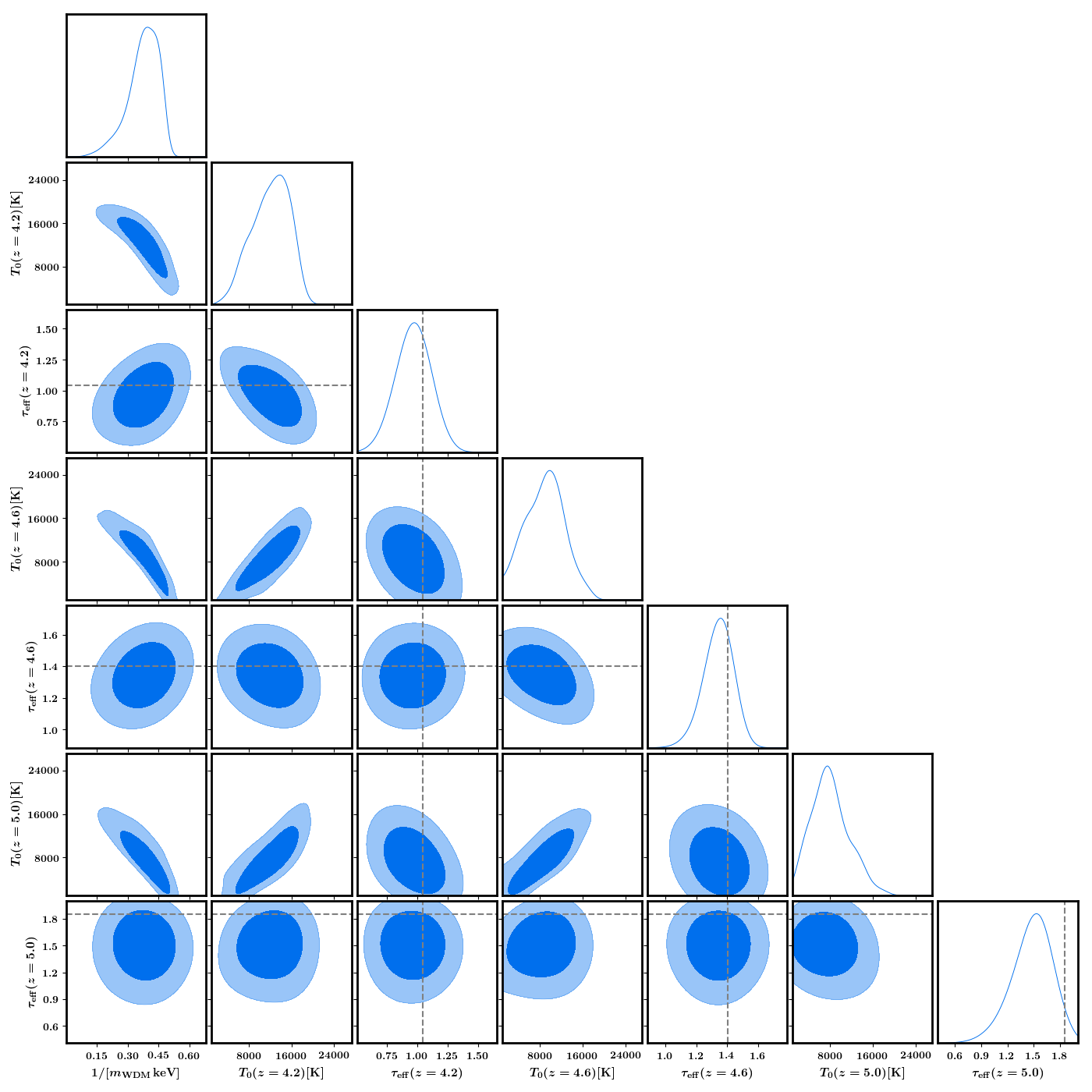}
  \caption{Confidence regions between the WDM mass, $m_\wdm$, the IGM mean
    temperature, $T_0$ and the effective optical depth, $\tau_{\rm eff}$, at redshifts $z=4.2, 4.6, 5.0$. The dashed grey lines are the
   $\tau_{\rm eff}$ as measured in \citep{Boera:2018vzq}.}
  \label{fig:triangle}
\end{figure*}

We explore the likelihood via the Monte Carlo Markov Chains
(MCMC). The priors are shown in Table~\ref{tab:priors}. The $1\sigma$ and $2\sigma$ contours of the cosmic
mean temperature, $T_0$, the optical depth, $\tau$, and the mass of
the WDM, $m_{\rm WDM}$ are shown in Figure~\ref{fig:2D}.
By marginalizing over $T_0(z)$ and $\tau(z)$ we find the lower limit on the warm dark matter $m_{\rm WDM}\geq 1.9\,{\rm keV}$ at $95\%$ CL.
Warmer WDM models are excluded regardless of the instantaneous temperature of IGM, as the contours in Fig.~\ref{fig:2D} demonstrate.  
For each mass we also find the \textit{upper bound} on $T_0(z)$ at redshifts $z=4.2, 4.6, 5.0$, see Fig.~\ref{fig:t0_imass_mcmc}.
A complete set of 2D plots are shown in Fig.~\ref{fig:triangle}.
We have repeated the MCMC exploration using the covariance matrix provided
by~\citet{Boera:2018vzq}, albeit with the errorbars inflated to account for
the sample variance. The resulting contours changed only slightly with the
$95\%$~CL for the mass reaching $2.05$~keV.

\begin{table}
\centering
\begin{tabular}{lcc}
\hline
parameter & lower range & upper range\\
\hline
$1/m_{\rm WDM}[1/{\rm keV}]$ & 0 & 1   \\
$T_0(z=4.2)[{\rm K}]$ & $10^3$ & $10^4$   \\
$T_0(z=4.6)[{\rm K}]$ & $10^3$ & $10^4$   \\
$T_0(z=5.0)[{\rm K}]$ & $10^3$ & $10^4$   \\
$\tau_{\rm eff}(z=4.2)$ & 0.5 & 2.0 \\
$\tau_{\rm eff}(z=4.6)$ & 0.5 & 2.0 \\
$\tau_{\rm eff}(z=5.0)$ & 0.5 & 2.0 \\
\hline
\end{tabular}
\caption{We show the priors that have been considered in our analysis.}
\label{tab:priors}
\end{table}

As our procedure is somewhat different from a standard MCMC exploration of the likelihood, we have several comments:
\begin{enumerate}
    \item Our contours do not reach the CDM values (${1/m_\wdm = 0}$).
This does not mean that the CDM cosmology is excluded by the data.  
This  indicates that in the {\sc LateCold} reionizaiton scenario with
cold DM, the temperature alone would not be sufficient to explain the
suppression of the FPS. For this reason, we only consider the upper
portion of the contours that are depicted  in Figure~\ref{fig:2D}.
Therefore, in CDM cosmology the {\sc LateCold} model would e ruled out, while in the WDM cosmology with $m_\wdm >\unit[1.9]{keV}$ it is actually allowed. 
Ref.~\cite{Walther:2018pnn} studied the dataset of \cite{Viel13} and  found that cold DM model with {\sc LateCold} reionization is consistent  with  the  data, but they did not study the dataset~\cite{Boera:2018vzq} we have considered in this work.

\item For the same reason our upper limits on the temperature (Fig.~\ref{fig:t0_imass_mcmc}) are, probably, exaggerated as they compensate minimal pressure history. While for the {\sc LateCold} history these are  $95\%$~CL upper bounds, for other thermal histories at similar confidence level one would be getting lower temperatures. Thus, one can treat them as (overly) conservative upper bounds on $T_0$. 
We do not provide lower bounds on the temperature as they are not meaningful.
\end{enumerate}

\section{Conclusion and future work}
\label{sec:conclusion}

\begin{figure}
    \centering
    \includegraphics[width=\linewidth]{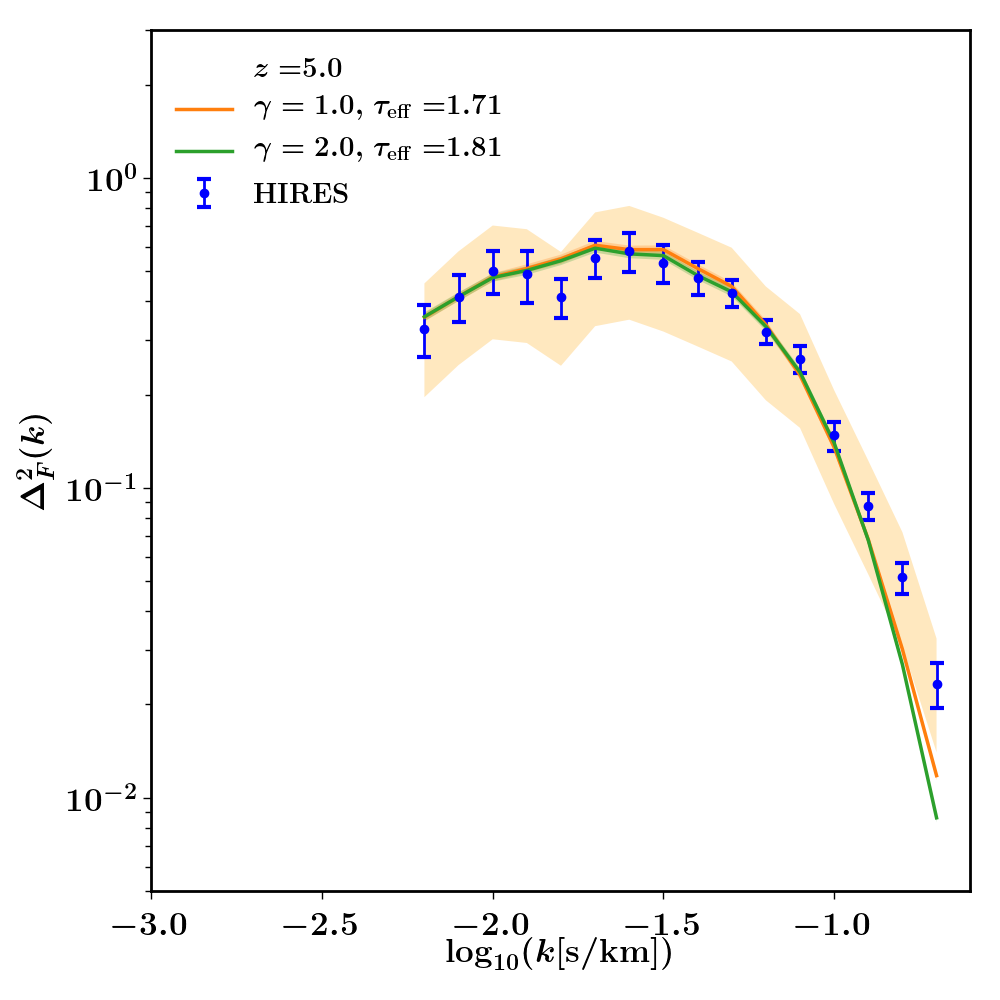}
    \caption{Dependence of the flux power spectrum on gamma can be ignored, given a poorly constrainted overall normalization $\teff$.}
    \label{fig:gamma}
\end{figure}
We have produced new constraint on the mass of warm dark matter particles using the high resolution Lyman-$\alpha$ forest dataset \citep{Boera:2018vzq}.
In order to minimize astrophysical effects and to leave the room for warm dark matter, we have used a conservative IGM thermal history that gives a minimal size of the intergalactic structures (the smallest pressure effects).
We have also explicitly marginalized over the Doppler broadening, finding that WDM thermal relics with mass as low as $ 1.9$~keV are consistent with the data at $95\%$~CL. 
Our  procedure lowers the Lyman-$\alpha$ bound significantly compared to $m_\wdm \ge  3.3$~keV obtained from the HIRES/MIKE data by~\cite{Viel13} (for detailed comparison between assumed thermal histories in these works see the discussion in~\cite{Garzilli:2015iwa,Garzilli18}).\footnote{All limits quoted at this section are at 95\% CL.}
In terms of the characteristic free-streaming length, $\lambda_\dm$ our bound is two times weaker.

Our result does not include marginalization over other  astrophysical ``nuisance'' parameters: a powerlaw index $\gamma$ in the temperature-density relation~\eqref{eq:TDR}, the AGN feedback, fluctuations of UVB, etc.
In Fig.~\ref{fig:gamma} we show that the difference between two highly distinct $\gamma$ values can be compensated by the different choice of $\teff$, both consistent with measurements. 
Other effects have been explored in other works \citep[see e.g.][]{Bertone:2005ud,Becker:2014oga,Garzilli:2015iwa,DAloisio:2016dcv,Chabanier:2018rga,Wu:2019sgk,Sanderbeck:2020vkn}. 
We stress that our goal was not to provide a most robust bound on WDM
mass, but rather to assess a level of systematic uncertainty as one
varies the pressure support scale. 

Because we have chosen a thermal history that minimize the cut-off on
the flux PS, our bound is independent of the reionization history.

Although the difference between CDM and WDM is the most pronounced at small scales, the influence of astrophysical effects is also the largest there.
Therefore, it becomes interesting to compare our results  with those, obtained from the medium resolution SDSS Lyman-$\alpha$ forest datasets.
These data have several advantages that may help to reduce this systematic uncertainty:  reduced sample variance due to the large number of quasars; less pronounced dependence on astrophysical processes on  larger scales.
Using the SDSS dataset,~\citet{Seljak06}  found $m_\wdm \gtrsim 2.5$~keV \citep[see also]{Viel06}, which ref.~\citet{Boyarsky09a} reduced to $m_\wdm \ge 1.7$~keV, using a more conservative treatment of the systematic uncertainties. 
The most recent analysis based on the SDSS-III/BOSS dataset~\cite{Baur16,baur2017} found  $m_\wdm \ge 3.06$~keV (when using Planck cosmological parameters). 
These works did not explore the influence of the thermal histories similar to {\sc LateCold} used here, as discussed in the Introduction.
We leave investigation of this question for a future work.

In this work we adopted {\sc LateCold} reinoization history as the one, providing minimal pressure support, consistent with existing data. 
Whether this is the case in non-CDM cosmologies (not explored in~\cite{onorbe2016}) or whether minimal pressure scale $\lambda_p$ can be estimated by other methods (\textit{e.g.}, \cite{Garzilli:2015bha,Garzilli:2018jna}) or from other data (see \textit{e.g.},~\cite{Telikova:2019eef}) remains to be seen.

\bigskip

We note that there are many other bounds on WDM, not based on the Lyman-$\alpha$ forest data (for example, those based on gravitational lensing \citep{Hsueh:2019ynk}, stellar streams \citep{Banik:2019smi}, satellite counts \citep{Nadler:2020prv}), each of them claiming bounds of $\lambda_\dm \lesssim \mathcal{O}(10-30)$~ckpc.
At these small scales the influence of baryonic physics is significant (even dominant) and each of these bounds suffer from their own systematic uncertainties.
Therefore, it is important to assess the systematic uncertainty of each of the methods. 
Our work makes the first step in this direction for the Lyman-$\alpha$ forest method.

\myparagraph{Acknowledgments.}  This project has received funding from the
European Research Council (ERC) under the European Union's Horizon
2020 research and innovation programme (ERC Advanced Grant 694896).
This work used the DiRAC Data Centric system at Durham University,
operated by the Institute for Computational Cosmology on behalf of the
STFC DiRAC HPC Facility (www.dirac.ac.uk). This equipment was funded
by BIS National E-infrastructure capital grant ST/K00042X/1, STFC
capital grants ST/H008519/1 and ST/K00087X/1, STFC DiRAC Operations
grant ST/K003267/1 and Durham University. DiRAC is part of the
National E-Infrastructure.
AM is supported by the Netherlands Organization for Scientific Research (NWO) under the program "Observing the Big Bang" of the Organization for Fundamental Research in Master (FOM).

We would like to thank Jose O{\~n}orbe for sharing with us additional
unpublished thermal histories.

\myparagraph{Data availability.}
The data that support the findings of this study are available from
the corresponding author, A.G., upon reasonable request.

\bibliography{Lyman_alpha_new_hires}

\begin{thebibliography}{}
\interlinepenalty=10000
\makeatletter
\relax
\def\mn@urlcharsother{\let\do\@makeother \do\$\do\&\do\#\do\^\do\_\do\%\do\~}
\def\mn@doi{\begingroup\mn@urlcharsother \@ifnextchar [ {\mn@doi@}
  {\mn@doi@[]}}
\def\mn@doi@[#1]#2{\def\@tempa{#1}\ifx\@tempa\@empty \href
  {http://dx.doi.org/#2} {doi:#2}\else \href {http://dx.doi.org/#2} {#1}\fi
  \endgroup}
\def\mn@eprint#1#2{\mn@eprint@#1:#2::\@nil}
\def\mn@eprint@arXiv#1{\href {http://arxiv.org/abs/#1} {{\tt arXiv:#1}}}
\def\mn@eprint@dblp#1{\href {http://dblp.uni-trier.de/rec/bibtex/#1.xml}
  {dblp:#1}}
\def\mn@eprint@#1:#2:#3:#4\@nil{\def\@tempa {#1}\def\@tempb {#2}\def\@tempc
  {#3}\ifx \@tempc \@empty \let \@tempc \@tempb \let \@tempb \@tempa \fi \ifx
  \@tempb \@empty \def\@tempb {arXiv}\fi \@ifundefined
  {mn@eprint@\@tempb}{\@tempb:\@tempc}{\expandafter \expandafter \csname
  mn@eprint@\@tempb\endcsname \expandafter{\@tempc}}}

\bibitem[\protect\citeauthoryear{Ade et~al.}{Ade et~al.}{2016}]{Planck2015}
Ade P. A.~R.,  et~al., 2016, \mn@doi [Astron. Astrophys.]
  {10.1051/0004-6361/201525830}, 594, A13

\bibitem[\protect\citeauthoryear{Banik, Bovy, Bertone, Erkal  \& de Boer}{Banik
  et~al.}{2019}]{Banik:2019smi}
Banik N.,  Bovy J.,  Bertone G.,  Erkal D.,   de Boer T. J.~L.,  2019

\bibitem[\protect\citeauthoryear{{Baur}, {Palanque-Delabrouille}, {Yeche},
  {Magneville}  \& {Viel}}{{Baur} et~al.}{2016}]{Baur16}
{Baur} J.,  {Palanque-Delabrouille} N.,  {Yeche} C.,  {Magneville} C.,   {Viel}
  M.,  2016, \mn@doi [\jcap] {10.1088/1475-7516/2016/08/012}, \href
  {http://adsabs.harvard.edu/abs/2016JCAP...08..012B} {8, 012}

\bibitem[\protect\citeauthoryear{{Baur}, {Palanque-Delabrouille}, {Y{\`e}che},
  {Boyarsky}, {Ruchayskiy}, {Armengaud}  \& {Lesgourgues}}{{Baur}
  et~al.}{2017}]{baur2017}
{Baur} J.,  {Palanque-Delabrouille} N.,  {Y{\`e}che} C.,  {Boyarsky} A.,
  {Ruchayskiy} O.,  {Armengaud} {\'E}.,   {Lesgourgues} J.,  2017, \mn@doi
  [\jcap] {10.1088/1475-7516/2017/12/013}, \href
  {http://adsabs.harvard.edu/abs/2017JCAP...12..013B} {12, 013}

\bibitem[\protect\citeauthoryear{Becker et~al.}{Becker
  et~al.}{2001}]{Becker:2001ee}
Becker R.~H.,  et~al., 2001, \mn@doi [Astron. J.] {10.1086/324231}, 122, 2850

\bibitem[\protect\citeauthoryear{Becker, Rauch  \& Sargent}{Becker
  et~al.}{2007}]{becker2007}
Becker G.~D.,  Rauch M.,   Sargent W. L.~W.,  2007, \mn@doi [Astrophys. J.]
  {10.1086/517866}, 662, 72

\bibitem[\protect\citeauthoryear{{Becker}, {Bolton}, {Haehnelt}  \&
  {Sargent}}{{Becker} et~al.}{2011}]{becker2011}
{Becker} G.~D.,  {Bolton} J.~S.,  {Haehnelt} M.~G.,   {Sargent} W.~L.~W.,
  2011, \mn@doi [\mnras] {10.1111/j.1365-2966.2010.17507.x}, \href
  {http://adsabs.harvard.edu/abs/2011MNRAS.410.1096B} {410, 1096}

\bibitem[\protect\citeauthoryear{Becker, Bolton, Madau, Pettini, Ryan-Weber  \&
  Venemans}{Becker et~al.}{2015}]{Becker:2014oga}
Becker G.~D.,  Bolton J.~S.,  Madau P.,  Pettini M.,  Ryan-Weber E.~V.,
  Venemans B.~P.,  2015, \mn@doi [Mon. Not. Roy. Astron. Soc.]
  {10.1093/mnras/stu2646}, 447, 3402

\bibitem[\protect\citeauthoryear{{Bernstein}, {Athey}, {Bernstein}, {Gunnels},
  {Richstone}  \& {Shectman}}{{Bernstein} et~al.}{2002}]{MIKE}
{Bernstein} G.~M.,  {Athey} A.~E.,  {Bernstein} R.,  {Gunnels} S.~M.,
  {Richstone} D.~O.,   {Shectman} S.~A.,  2002, {Volume-phase holographic
  spectrograph for the Magellan telescopes}.
pp 453--459, \mn@doi{10.1117/12.454281}

\bibitem[\protect\citeauthoryear{Bertone \& White}{Bertone \&
  White}{2006}]{Bertone:2005ud}
Bertone S.,  White S. D.~M.,  2006, \mn@doi [Mon. Not. Roy. Astron. Soc.]
  {10.1111/j.1365-2966.2005.09936.x}, 367, 247

\bibitem[\protect\citeauthoryear{Bode, Ostriker  \& Turok}{Bode
  et~al.}{2001}]{Bode:2000gq}
Bode P.,  Ostriker J.~P.,   Turok N.,  2001, \mn@doi [Astrophys. J.]
  {10.1086/321541}, 556, 93

\bibitem[\protect\citeauthoryear{Boera, Becker, Bolton  \& Nasir}{Boera
  et~al.}{2019}]{Boera:2018vzq}
Boera E.,  Becker G.~D.,  Bolton J.~S.,   Nasir F.,  2019, \mn@doi [Astrophys.
  J.] {10.3847/1538-4357/aafee4}, 872, 101

\bibitem[\protect\citeauthoryear{{Bolton}, {Becker}, {Wyithe}, {Haehnelt}  \&
  {Sargent}}{{Bolton} et~al.}{2010}]{bolton2010}
{Bolton} J.~S.,  {Becker} G.~D.,  {Wyithe} J.~S.~B.,  {Haehnelt} M.~G.,
  {Sargent} W.~L.~W.,  2010, \mn@doi [\mnras]
  {10.1111/j.1365-2966.2010.16701.x}, \href
  {http://adsabs.harvard.edu/abs/2010MNRAS.406..612B} {406, 612}

\bibitem[\protect\citeauthoryear{{Bolton}, {Becker}, {Raskutti}, {Wyithe},
  {Haehnelt}  \& {Sargent}}{{Bolton} et~al.}{2012}]{bolton2012}
{Bolton} J.~S.,  {Becker} G.~D.,  {Raskutti} S.,  {Wyithe} J.~S.~B.,
  {Haehnelt} M.~G.,   {Sargent} W.~L.~W.,  2012, \mn@doi [\mnras]
  {10.1111/j.1365-2966.2011.19929.x}, 419, 2880

\bibitem[\protect\citeauthoryear{{Boyarsky}, {Lesgourgues}, {Ruchayskiy}  \&
  {Viel}}{{Boyarsky} et~al.}{2009a}]{Boyarsky09a}
{Boyarsky} A.,  {Lesgourgues} J.,  {Ruchayskiy} O.,   {Viel} M.,  2009a,
  \mn@doi [\jcap] {10.1088/1475-7516/2009/05/012}, \href
  {http://adsabs.harvard.edu/abs/2009JCAP...05..012B} {5, 012}

\bibitem[\protect\citeauthoryear{Boyarsky, Lesgourgues, Ruchayskiy  \&
  Viel}{Boyarsky et~al.}{2009b}]{Boyarsky:2008mt}
Boyarsky A.,  Lesgourgues J.,  Ruchayskiy O.,   Viel M.,  2009b, \mn@doi [Phys.
  Rev. Lett.] {10.1103/PhysRevLett.102.201304}, 102, 201304

\bibitem[\protect\citeauthoryear{Chabanier et~al.}{Chabanier
  et~al.}{2019}]{Chabanier:2018rga}
Chabanier S.,  et~al., 2019, \mn@doi [\jcap] {10.1088/1475-7516/2019/07/017},
  1907, 017

\bibitem[\protect\citeauthoryear{D'Aloisio, McQuinn, Davies  \&
  Furlanetto}{D'Aloisio et~al.}{2018}]{DAloisio:2016dcv}
D'Aloisio A.,  McQuinn M.,  Davies F.~B.,   Furlanetto S.~R.,  2018, \mn@doi
  [Mon. Not. Roy. Astron. Soc.] {10.1093/mnras/stx2341}, 473, 560

\bibitem[\protect\citeauthoryear{Gaikwad et~al.}{Gaikwad
  et~al.}{2020}]{Gaikwad:2020art}
Gaikwad P.,  et~al., 2020, \mn@doi [Mon. Not. Roy. Astron. Soc.]
  {10.1093/mnras/staa907}, 494, 5091

\bibitem[\protect\citeauthoryear{Garzilli, Theuns  \& Schaye}{Garzilli
  et~al.}{2015}]{Garzilli:2015bha}
Garzilli A.,  Theuns T.,   Schaye J.,  2015, \mn@doi [Mon. Not. Roy. Astron.
  Soc.] {10.1093/mnras/stv394}, 450, 1465

\bibitem[\protect\citeauthoryear{Garzilli, Boyarsky  \& Ruchayskiy}{Garzilli
  et~al.}{2017}]{Garzilli:2015iwa}
Garzilli A.,  Boyarsky A.,   Ruchayskiy O.,  2017, \mn@doi [Phys. Lett.]
  {10.1016/j.physletb.2017.08.022}, B773, 258

\bibitem[\protect\citeauthoryear{{Garzilli}, {Magalich}, {Theuns}, {Frenk},
  {Weniger}, {Ruchayskiy}  \& {Boyarsky}}{{Garzilli} et~al.}{2019}]{Garzilli18}
{Garzilli} A.,  {Magalich} A.,  {Theuns} T.,  {Frenk} C.~S.,  {Weniger} C.,
  {Ruchayskiy} O.,   {Boyarsky} A.,  2019, \mn@doi [\mnras]
  {10.1093/mnras/stz2188}, \href
  {https://ui.adsabs.harvard.edu/abs/2019MNRAS.tmp.2103G} {p.~2103}

\bibitem[\protect\citeauthoryear{Garzilli, Theuns  \& Schaye}{Garzilli
  et~al.}{2020}]{Garzilli:2018jna}
Garzilli A.,  Theuns T.,   Schaye J.,  2020, \mn@doi [Mon. Not. Roy. Astron.
  Soc.] {10.1093/mnras/stz3585}, 492, 2193

\bibitem[\protect\citeauthoryear{{Gnedin} \& {Hui}}{{Gnedin} \&
  {Hui}}{1998}]{Gnedin98}
{Gnedin} N.~Y.,  {Hui} L.,  1998, \mn@doi [\mnras]
  {10.1046/j.1365-8711.1998.01249.x}, \href
  {http://adsabs.harvard.edu/abs/1998MNRAS.296...44G} {296, 44}

\bibitem[\protect\citeauthoryear{{Haardt} \& {Madau}}{{Haardt} \&
  {Madau}}{2001}]{haardt2001}
{Haardt} F.,  {Madau} P.,  2001, in {Neumann} D.~M.,  {Tran} J.~T.~V.,  eds,
  Clusters of Galaxies and the High Redshift Universe Observed in X-rays. ArXiv
  astro-ph/0106018 (\mn@eprint {} {arXiv:astro-ph/0106018})

\bibitem[\protect\citeauthoryear{{Haardt} \& {Madau}}{{Haardt} \&
  {Madau}}{2012}]{Haardt12}
{Haardt} F.,  {Madau} P.,  2012, \mn@doi [\apj] {10.1088/0004-637X/746/2/125},
  \href {http://adsabs.harvard.edu/abs/2012ApJ...746..125H} {746, 125}

\bibitem[\protect\citeauthoryear{Hansen, Lesgourgues, Pastor  \& Silk}{Hansen
  et~al.}{2002}]{Hansen01}
Hansen S.~H.,  Lesgourgues J.,  Pastor S.,   Silk J.,  2002, \mn@doi [Mon. Not.
  Roy. Astron. Soc.] {10.1046/j.1365-8711.2002.05410.x}, 333, 544

\bibitem[\protect\citeauthoryear{Hsueh, Enzi, Vegetti, Auger, Fassnacht,
  Despali, Koopmans  \& McKean}{Hsueh et~al.}{2020}]{Hsueh:2019ynk}
Hsueh J.-W.,  Enzi W.,  Vegetti S.,  Auger M.,  Fassnacht C.~D.,  Despali G.,
  Koopmans L. V.~E.,   McKean J.~P.,  2020, \mn@doi [Mon. Not. Roy. Astron.
  Soc.] {10.1093/mnras/stz3177}, 492, 3047

\bibitem[\protect\citeauthoryear{{Hui} \& {Haiman}}{{Hui} \&
  {Haiman}}{2003}]{hui2003}
{Hui} L.,  {Haiman} Z.,  2003, \mn@doi [\apj] {10.1086/377229}, 596, 9

\bibitem[\protect\citeauthoryear{{Hui} \& {Rutledge}}{{Hui} \&
  {Rutledge}}{1999}]{hui1999}
{Hui} L.,  {Rutledge} R.~E.,  1999, \mn@doi [\apj] {10.1086/307202}, \href
  {http://adsabs.harvard.edu/abs/1999ApJ...517..541H} {517, 541}

\bibitem[\protect\citeauthoryear{{Ir{\v s}i{\v c}} et~al.,}{{Ir{\v s}i{\v c}}
  et~al.}{2017}]{irsic2017}
{Ir{\v s}i{\v c}} V.,  et~al., 2017, \mn@doi [\prd]
  {10.1103/PhysRevD.96.023522}, \href
  {http://adsabs.harvard.edu/abs/2017PhRvD..96b3522I} {96, 023522}

\bibitem[\protect\citeauthoryear{Kulkarni, Hennawi, Oñorbe, Rorai  \&
  Springel}{Kulkarni et~al.}{2015}]{Kulkarni:2015fga}
Kulkarni G.,  Hennawi J.~F.,  Oñorbe J.,  Rorai A.,   Springel V.,  2015,
  \mn@doi [Astrophys. J.] {10.1088/0004-637X/812/1/30}, 812, 30

\bibitem[\protect\citeauthoryear{Lidz \& Malloy}{Lidz \&
  Malloy}{2014}]{Lidz:2014jxa}
Lidz A.,  Malloy M.,  2014, \mn@doi [Astrophys. J.]
  {10.1088/0004-637X/788/2/175}, 788, 175

\bibitem[\protect\citeauthoryear{{Meiksin}}{{Meiksin}}{2009}]{meiksin2009}
{Meiksin} A.~A.,  2009, \mn@doi [Rev. Mod. Phys.] {10.1103/RevModPhys.81.1405},
  \href {http://adsabs.harvard.edu/abs/2009RvMP...81.1405M} {81, 1405}

\bibitem[\protect\citeauthoryear{Murgia, {Ir{\v s}i{\v c}}  \& Viel}{Murgia
  et~al.}{2018}]{Murgia:2018now}
Murgia R.,  {Ir{\v s}i{\v c}} V.,   Viel M.,  2018, ArXiv preprint 1806.08371

\bibitem[\protect\citeauthoryear{Nadler et~al.}{Nadler
  et~al.}{2020}]{Nadler:2020prv}
Nadler E.~O.,  et~al., 2020

\bibitem[\protect\citeauthoryear{{O{\~n}orbe}, {Hennawi}  \&
  {Luki{\'c}}}{{O{\~n}orbe} et~al.}{2017a}]{onorbe2016}
{O{\~n}orbe} J.,  {Hennawi} J.~F.,   {Luki{\'c}} Z.,  2017a, \mn@doi [\apj]
  {10.3847/1538-4357/aa6031}, \href
  {http://adsabs.harvard.edu/abs/2017ApJ...837..106O} {837, 106}

\bibitem[\protect\citeauthoryear{{O{\~n}orbe}, {Hennawi}, {Luki{\'c}}  \&
  {Walther}}{{O{\~n}orbe} et~al.}{2017b}]{onorbe2017}
{O{\~n}orbe} J.,  {Hennawi} J.~F.,  {Luki{\'c}} Z.,   {Walther} M.,  2017b,
  \mn@doi [\apj] {10.3847/1538-4357/aa898d}, \href
  {http://adsabs.harvard.edu/abs/2017ApJ...847...63O} {847, 63}

\bibitem[\protect\citeauthoryear{Palanque-Delabrouille, Yèche, Schöneberg,
  Lesgourgues, Walther, Chabanier  \& Armengaud}{Palanque-Delabrouille
  et~al.}{2020}]{Palanque-Delabrouille:2019iyz}
Palanque-Delabrouille N.,  Yèche C.,  Schöneberg N.,  Lesgourgues J.,
  Walther M.,  Chabanier S.,   Armengaud E.,  2020, \mn@doi [JCAP]
  {10.1088/1475-7516/2020/04/038}, 2004, 038

\bibitem[\protect\citeauthoryear{Peebles}{Peebles}{2017}]{Peebles:2017bzw}
Peebles P. J.~E.,  2017, \mn@doi [Nat. Astron.] {10.1038/s41550-017-0057}, 1,
  0057

\bibitem[\protect\citeauthoryear{{Rollinde}, {Theuns}, {Schaye}, {P{\^a}ris}
  \& {Petitjean}}{{Rollinde} et~al.}{2013}]{rollinde2013}
{Rollinde} E.,  {Theuns} T.,  {Schaye} J.,  {P{\^a}ris} I.,   {Petitjean} P.,
  2013, \mn@doi [\mnras] {10.1093/mnras/sts057}, 428, 540

\bibitem[\protect\citeauthoryear{{Rorai}, {Carswell}, {Haehnelt}, {Becker},
  {Bolton}  \& {Murphy}}{{Rorai} et~al.}{2018}]{Rorai18}
{Rorai} A.,  {Carswell} R.~F.,  {Haehnelt} M.~G.,  {Becker} G.~D.,  {Bolton}
  J.~S.,   {Murphy} M.~T.,  2018, \mn@doi [\mnras] {10.1093/mnras/stx2862},
  \href {http://adsabs.harvard.edu/abs/2018MNRAS.474.2871R} {474, 2871}

\bibitem[\protect\citeauthoryear{Sanderbeck \& Bird}{Sanderbeck \&
  Bird}{2020}]{Sanderbeck:2020vkn}
Sanderbeck P.~U.,  Bird S.,  2020, ] {10.1093/mnras/staa1850}

\bibitem[\protect\citeauthoryear{{Schaye}, {Theuns}, {Rauch}, {Efstathiou}  \&
  {Sargent}}{{Schaye} et~al.}{2000}]{Schaye00}
{Schaye} J.,  {Theuns} T.,  {Rauch} M.,  {Efstathiou} G.,   {Sargent} W.~L.~W.,
   2000, \mn@doi [\mnras] {10.1046/j.1365-8711.2000.03815.x}, \href
  {http://adsabs.harvard.edu/abs/2000MNRAS.318..817S} {318, 817}

\bibitem[\protect\citeauthoryear{{Schaye}, {Crain}, {Bower}  \& al.}{{Schaye}
  et~al.}{2015}]{schaye2015}
{Schaye} J.,  {Crain} R.~A.,  {Bower} R.~G.,   al. 2015, \mn@doi [\mnras]
  {10.1093/mnras/stu2058}, \href
  {http://adsabs.harvard.edu/abs/2015MNRAS.446..521S} {446, 521}

\bibitem[\protect\citeauthoryear{{Schroeder}, {Mesinger}  \&
  {Haiman}}{{Schroeder} et~al.}{2013}]{schroeder2013}
{Schroeder} J.,  {Mesinger} A.,   {Haiman} Z.,  2013, \mn@doi [\mnras]
  {10.1093/mnras/sts253}, \href
  {http://adsabs.harvard.edu/abs/2013MNRAS.428.3058S} {428, 3058}

\bibitem[\protect\citeauthoryear{{Scoccimarro}, {Hui}, {Manera}  \&
  {Chan}}{{Scoccimarro} et~al.}{2012}]{scoccimarro2012}
{Scoccimarro} R.,  {Hui} L.,  {Manera} M.,   {Chan} K.~C.,  2012, \mn@doi
  [\prd] {10.1103/PhysRevD.85.083002}, \href
  {http://adsabs.harvard.edu/abs/2012PhRvD..85h3002S} {85, 083002}

\bibitem[\protect\citeauthoryear{Seljak, Makarov, McDonald  \& Trac}{Seljak
  et~al.}{2006}]{Seljak06}
Seljak U.,  Makarov A.,  McDonald P.,   Trac H.,  2006, \mn@doi [Phys. Rev.
  Lett.] {10.1103/PhysRevLett.97.191303}, 97, 191303

\bibitem[\protect\citeauthoryear{{Springel}}{{Springel}}{2005}]{springel2005}
{Springel} V.,  2005, \mn@doi [\mnras] {10.1111/j.1365-2966.2005.09655.x},
  \href {http://adsabs.harvard.edu/abs/2005MNRAS.364.1105S} {364, 1105}

\bibitem[\protect\citeauthoryear{Telikova, Balashev  \& Shternin}{Telikova
  et~al.}{2019}]{Telikova:2019eef}
Telikova K.~N.,  Balashev S.~A.,   Shternin P.~S.,  2019, \mn@doi [J. Phys.
  Conf. Ser.] {10.1088/1742-6596/1400/2/022024}, 1400, 022024

\bibitem[\protect\citeauthoryear{{Theuns}, {Schaye}  \& {Haehnelt}}{{Theuns}
  et~al.}{2000}]{Theuns00}
{Theuns} T.,  {Schaye} J.,   {Haehnelt} M.~G.,  2000, \mn@doi [\mnras]
  {10.1046/j.1365-8711.2000.03423.x}, \href
  {http://adsabs.harvard.edu/abs/2000MNRAS.315..600T} {315, 600}

\bibitem[\protect\citeauthoryear{{Upton Sanderbeck}, {D'Aloisio}  \&
  {McQuinn}}{{Upton Sanderbeck} et~al.}{2016}]{Upton16}
{Upton Sanderbeck} P.~R.,  {D'Aloisio} A.,   {McQuinn} M.~J.,  2016, \mn@doi
  [\mnras] {10.1093/mnras/stw1117}, \href
  {http://adsabs.harvard.edu/abs/2016MNRAS.460.1885U} {460, 1885}

\bibitem[\protect\citeauthoryear{Viel, Lesgourgues, Haehnelt, Matarrese  \&
  Riotto}{Viel et~al.}{2005}]{Viel05}
Viel M.,  Lesgourgues J.,  Haehnelt M.~G.,  Matarrese S.,   Riotto A.,  2005,
  \mn@doi [Phys. Rev.] {10.1103/PhysRevD.71.063534}, D71, 063534

\bibitem[\protect\citeauthoryear{Viel, Lesgourgues, Haehnelt, Matarrese  \&
  Riotto}{Viel et~al.}{2006}]{Viel06}
Viel M.,  Lesgourgues J.,  Haehnelt M.~G.,  Matarrese S.,   Riotto A.,  2006,
  \mn@doi [Phys. Rev. Lett.] {10.1103/PhysRevLett.97.071301}, 97, 071301

\bibitem[\protect\citeauthoryear{Viel, Becker, Bolton, Haehnelt, Rauch  \&
  Sargent}{Viel et~al.}{2008}]{Viel:2007mv}
Viel M.,  Becker G.~D.,  Bolton J.~S.,  Haehnelt M.~G.,  Rauch M.,   Sargent W.
  L.~W.,  2008, \mn@doi [Phys. Rev. Lett.] {10.1103/PhysRevLett.100.041304},
  100, 041304

\bibitem[\protect\citeauthoryear{{Viel}, {Becker}, {Bolton}  \&
  {Haehnelt}}{{Viel} et~al.}{2013}]{Viel13}
{Viel} M.,  {Becker} G.~D.,  {Bolton} J.~S.,   {Haehnelt} M.~G.,  2013, \mn@doi
  [\prd] {10.1103/PhysRevD.88.043502}, \href
  {http://adsabs.harvard.edu/abs/2013PhRvD..88d3502V} {88, 043502}

\bibitem[\protect\citeauthoryear{Vogt et~al.}{Vogt et~al.}{1994}]{Vogt:1995zz}
Vogt S.~S.,  et~al., 1994, \mn@doi [Proc. SPIE Int. Soc. Opt. Eng.]
  {10.1117/12.176725}, 2198, 362

\bibitem[\protect\citeauthoryear{Walther, Hennawi, Hiss, Oñorbe, Lee, Rorai
  \& O'Meara}{Walther et~al.}{2018}]{Walther:2017cir}
Walther M.,  Hennawi J.~F.,  Hiss H.,  Oñorbe J.,  Lee K.-G.,  Rorai A.,
  O'Meara J.,  2018, \mn@doi [Astrophys. J.] {10.3847/1538-4357/aa9c81}, 852,
  22

\bibitem[\protect\citeauthoryear{Walther, O{\~n}orbe, Hennawi  \&
  Luki\'c}{Walther et~al.}{2019}]{Walther:2018pnn}
Walther M.,  O{\~n}orbe J.,  Hennawi J.~F.,   Luki\'c Z.,  2019, \mn@doi
  [Astrophys. J.] {10.3847/1538-4357/aafad1}, 872, 13

\bibitem[\protect\citeauthoryear{Wu, McQuinn, Kannan, D'Aloisio, Bird,
  Marinacci, Davé  \& Hernquist}{Wu et~al.}{2019}]{Wu:2019sgk}
Wu X.,  McQuinn M.,  Kannan R.,  D'Aloisio A.,  Bird S.,  Marinacci F.,  Davé
  R.,   Hernquist L.,  2019, \mn@doi [Mon. Not. Roy. Astron. Soc.]
  {10.1093/mnras/stz2807}, 490, 3177

\makeatother
\end{thebibliography}

\label{lastpage}

\end{document}